# Metasurface-controlled holographic microcavities

Sydney Mason[1, †], Maryna Leonidivna Meretska[1, ‡], Christina Spägele[1], Marcus Ossiander[1, 2, *], Federico Capasso[1]

[1] John A. Paulson School of Engineering and Applied Sciences, Harvard University, Cambridge, MA 02138, USA

[2] Institute of Experimental Physics, Graz University of Technology, 8010 Graz, Austria

**ABSTRACT:**

Optical microcavities confine light to wavelength-scale volumes and are a key component for manipulating and enhancing the interaction of light, vacuum states, and matter. Current microcavities are constrained to a small number of spatial mode profiles. Imaging cavities can accommodate complicated modes but require an externally pre-shaped input. Here, we experimentally demonstrate a visible-wavelength, metasurface-based, holographic microcavity that overcomes these limitations. The micron-scale metasurface cavity fulfills the round-trip condition for a designed mode with a complex-shaped intensity profile and thus selectively enhances light that couples to this mode, achieving a spectral bandwidth of 0.8 nm. By imaging the intracavity mode, we show that the holographic mode changes quickly with the cavity length, and the cavity displays the desired spatial mode profile only close to the design cavity length. When placing a metasurface on a distributed Bragg reflector and realizing steep phase gradients, the correct choice of the reflector's top layer material can boost metasurface performance considerably. The applied forward-design method is readily transferable to other spectral regimes and mode profiles.

**KEYWORDS:** metaoptics, photonic cavity, hologram, optical metamaterials, mode shaping

## INTRODUCTION

Optical cavities trap light and are useful for applications in linear, nonlinear, and quantum optics, metrology, communications, and computing[1]. Micro-scale cavities, with mode volumes on the order of a few wavelengths cubed, additionally provide precise control over the state and evolution of matter within them[2,3,4]. For example, cavities for the detection of molecules[5] and the imaging of nanoparticles[6] have been developed, and cavity quantum electrodynamics has been exploited to trap and manipulate single atoms[7]. This makes optical cavities good candidates for applications in quantum networks and nanoscale sensing[8]. So far, most cavities are designed using conventional optics and thus support a limited number of simple spatial mode profiles[9,10]. Special cavity geometries can exhibit imaging properties[11,12,13], which provide field enhancement for designed mode profiles. However, their size is on the order of millimeters, and they require the external injection of the desired mode profile[13].

Optical metasurfaces allow control of the intensity, polarization, and phase of light on the subwavelength scale[14,15,16] and can be used, e.g., as polarizers, modulators, surface wave absorbers, waveguides[15], and objective lenses[17]. High-fidelity computer-generated holography is another important metasurface use case[18,19,20] with applications in augmented reality[21], security, and encryption[22]. Metasurfaces have so far been used to counteract the transverse mode expansion in cavities and such to provide stable microcavities without concave mirrors at microwave frequencies[23], at telecom wavelengths[24], for external cavity lasers[19], and within hollow-core fibers[25] but not in the visible spectrum with free-space cavities. Additionally, the large phase response and spectral control provided by metasurfaces have allowed for nanoscale cavities on the order of a metaatom[26,27]. Metasurface cavities have been designed to achieve a near-zero index allowing for on-chip phase matching[26]. This approach can be used to enhance integrated photonic platforms. Other approaches to metasurface optical cavities exploit metasurfaces for decreasing the cavity size[27], steering the mode exiting the cavity[28,29], and altering the cavity spectrum[27,30,31].

This selection demonstrates the impressive versatility of metasurface microcavities and their multifarious applications, many of which accept or even require a broadband cavity response. However, other applications, e.g., field enhancement, demand cavities with narrow spectral resonances. To achieve a cavity mode with a narrow linewidth, its transverse phase evolution must be carefully controlled every time light propagates a full round-trip in the cavity.

Recently, two works have shown such transverse phase control using metasurfaces and have experimentally demonstrated stable Gaussian modes in microcavities[23,24]. For Gaussian-shaped modes, the metasurface approaches have not yet exceeded the capabilities of state-of-the-art microscale spherical mirrors[9]. However, metasurfaces can realize considerably steeper phase gradients than microscale mirrors which are limited by cracks that form during their fabrication if their topography changes quickly[9].

Due to their ability for sub-wavelength phase control, metasurfaces have been predicted to allow intracavity microscale holography[24], i.e., stable cavity modes with a complicated spatial profile. So far, the idea has only been theoretically explored for infrared light. However, the approach has not yet been experimentally demonstrated.

Here we fill this gap and experimentally create an open metasurface microcavity operating in the visible (λ = 633 nm). Through a computer-generated holography design approach, the miniaturized optical cavity creates a complex-shaped intracavity mode that classical optics cannot achieve. Its spatial profile can be non-spherical, asymmetrical, and can comprise multiple narrow or differing features.

While previous work on metasurface optical microcavities[27,30,31] has exploited metasurfaces to control the spectral characteristics of cavity resonances and to shape the output of cavities[19,28,29], here we for the first time experimentally realize a stable cavity mode with a complicated spatial profile inside the cavity. We image the cavity-length dependent microscale intracavity mode experimentally and demonstrate the concept synthesizes the designed transverse mode at a specific cavity length.

## METHOD

The setup of our metasurface microcavity is shown in Fig. 1a: it consists of two opposing, flat partial reflectors with a metasurface placed on one of the reflectors. This setup resembles a plano-concave cavity, with the metasurface taking the role of the spherical mirror. As indicated in Fig. 1a, from now on we will label the top of the uncovered partial reflector (PR) as the image plane (IP), and the top of the metasurface-covered reflector (MS) as the metasurface plane (MP).

For a mode to build up in the cavity, the phase and intensity profiles of the mode must reproduce after each round-trip, i.e., for each transverse position $(x, y)$ in the metasurface plane, the round-trip phase $\Delta\phi(x, y)$ accrued by the light within the cavity is an integer multiple $q$ of $2\pi$. The contributions to $\Delta\phi$ are the mode's propagation phase from the image plane to the metasurface plane $\phi_{IP \to MP}^{forward}$, the spatially dependent metasurface reflection phase $\phi_{MS}$, the mode's propagation phase back to the image plane $\phi_{MP \to IP}^{forward}$, and the spatially independent reflection phase of the partial reflector $\phi_{PR}$:

$$\Delta\phi(x, y) = 2\pi q = \phi_{IP \to MP}^{forward} + \phi_{MS} + \phi_{MP \to IP}^{forward} + \phi_{PR} \tag{1}$$

When we explicitly include the spatial dependence, all phase contributions must be treated in a single plane, i.e., the metasurface plane. Therefore, we replace the forward propagation phase from the metasurface plane to the image plane $\phi_{MP \to IP}^{forward}(x', y')$ with $-\phi_{IP \to MP}^{backward}(x, y)$. A full round-trip is then described by:

$$\Delta\phi(x, y) = 2\pi q = \phi_{IP \to MP}^{forward}(x, y) + \phi_{MS}(x, y) - \phi_{IP \to MP}^{backward}(x, y) + \phi_{PR} \tag{2}$$

The propagation phases also depend on the length of the cavity $L_{cav}$, indicated in Fig. 1a.

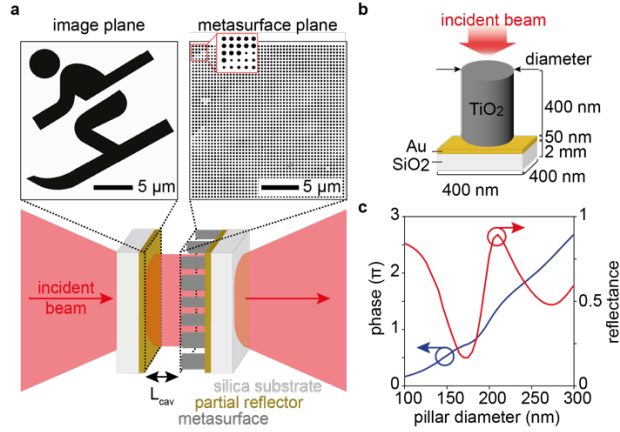

*Figure 1. Holographic microcavity concept and design. a) Holographic microcavity that forms an intracavity image: unstructured light is incident from the left. In the cavity, a metasurface (right inset) introduces a reflective phase profile that fulfills the round-trip phase resonance condition for a designed mode. Because light only interferes constructively if the round-trip condition is met, only the mode with the desired spatial intensity profile in the image plane (left inset) is excited in the cavity. The desired intensity profile is encoded in the metasurface as a phase-only hologram by suitable design of the nanopillars' phase shifts. For this demonstration, we chose a spatial mode profile resembling a skier. b) Metaatom schematic used to create the nanopillar library in panel c. Light is incident normally on a circular titania ($TiO_2$) nanopillar which sits on a partially reflective gold (Au) layer on a thick fused silica ($SiO_2$) substrate. c) Nanopillar library for visible light (633 nm). Whereas the gold layer controls the average reflectance (red line), the nanopillars' reflection phase (blue line) can be tuned by more than $2\pi$ via their diameter.*

**Metasurface Phase Calculation**

To create a microcavity that selectively enhances a mode with a spatial intensity profile $I_{IP}(x', y')$ in the image plane $x', y'$, we design a metasurface that fulfills the round-trip condition for this mode profile.

Assuming a flat phase in the imaging plane, the complex electric field amplitude of the mode is proportional to the square root of its intensity $E_{IP}(x', y') \sim \sqrt{I_{IP}(x', y')}$. We can then calculate the evolution of this mode along the propagation direction $z$ using the Rayleigh-Sommerfeld diffraction integral[24]:

$$E(x, y, z, k) = -\frac{ik}{2\pi} \int \int dx' dy' E_{IP}(x', y') \frac{e^{\pm ikr}}{r} cos(\chi) \tag{3}$$

$$r = \sqrt{(x - x')^2 + (y - y')^2 + z^2} \tag{4}$$

$$\chi = tan^{-1}\left(\frac{\sqrt{(x-x')^2+(y-y')^2}}{z}\right) \tag{5}$$

The equation describes how the metasurface is designed: we search for a metasurface phase that fulfills the round-trip condition for an already known image $I_{IP}(x', y')$. In eq. (3), we use the positive propagator for forward propagation and the negative propagator for backward propagation. The phase difference between the forward-propagated mode field $E_{IP \to MP}^{forward}(x, y)$ and the backward-propagated mode field $E_{IP \to MP}^{backward}(x, y)$ then represents the round-trip propagation phase without the reflector phases (see eq. (2)):

$$\phi_{IP \to MP}^{forward}(x, y) - \phi_{IP \to MP}^{backward}(x, y) = arg\left(E_{IP \to MP}^{forward}(x, y)\right) - arg\left(E_{IP \to MP}^{backward}(x, y)\right) \tag{6}$$

Consequently, we can fulfill the round-trip condition eq. (2) for a designer mode with an intensity profile $I_{IP}(x', y')$ with a metasurface that realizes a spatial phase profile:

$$\phi_{MS}(x, y) = 2\pi q - arg\left(E_{IP \to MP}^{forward}(x, y)\right) + arg(E_{IP \to MP}^{backward}(x, y)) - \phi_{PR} \tag{7}$$

Because $\phi_{PR}$ is spatially independent, we neglect it in the following.

To demonstrate this design concept's capability for adapting an arbitrary mode profile, we chose a skier shape (Fig. 1a inset) as the designed mode profile. The shape exhibits small features with bars as narrow as 1.5 μm. Furthermore, we chose an operating wavelength of 633 nm, a cavity mirror reflectance of 50 %, a metasurface size of 20 μm x 20 μm, and a cavity length of 40 μm. The finite transverse size limits the effective numerical aperture of our cavity, i.e., the maximum transverse momentum components of light that can be trapped by it. This, in turn, limits the steepness of the intensity gradients realizable in such a cavity. The effect of the finite cavity size is illustrated in Fig. 4a, which still shows the desired mode with high fidelity.

**Metasurface Cavity Design**

Our metaatom schematic is pictured in Fig. 1b: as this demonstration does not require the high reflectances of distributed Bragg reflectors (DBRs), which are typically used to realize cavities with high finesses[32], we used metallic partial reflectors, which allow easy tuning of the reflectance via the metal thickness. We use a 50-nm thick layer of gold on the fused silica substrates for the cavity-end reflectors, which resulted in 46 % reflectance at 633 nm. We chose titania nanopillars as metaatoms because of titania's low absorption and high refractive index in the visible[33]. We then created a reflectance library of metaatoms through finite-difference time-domain modeling (Lumerical FDTD) of nanopillars with a constant height (400 nm), a constant periodicity (400 nm x 400 nm), and diameters ranging from 100-300 nm. The resulting library is shown in Fig. 1c and allows changes of the reflection phase by more than $2\pi$ with an average 57 % reflectance across the nanopillar distribution.

We calculate the evolution of the skier mode within the cavity using eq. (3) and discretize the phase resulting from eq. (7) into 400 nm x 400 nm cells, see Fig. 2a. Metaatoms with the corresponding phases are selected from the library and mapped to each cell to build the metasurface design. The inset in Fig. 1a shows a top view of the final metasurface design.

The final performance of the metasurface cavity depends on the wavelength, the cavity length and transverse size, the employed materials and reflectors, fabrication quality, and principally on the complexity of the desired mode profile. Extreme gradients in the pillar diameter, caused, e.g., by very narrow features in the mode profile or high numerical apertures, impact the holographic cavity's performance. Carefully selecting the metaatoms such that their pillar sizes change slowly in the center of the metasurface improves performance[24]. Although the choice of the mode profile for this demonstration, the skier, is subject to such limitations, we show the robustness of this design platform by choosing a complex-shaped spatial mode. In practice, the performance of the device will to some extend depend on the choice of the hologram and application.

Contrary to common holography algorithms[34], designing cavity holograms using the resonance condition (eq. (1)) requires no iteration. The design flow is as follows:

choose the spatial profile of the desired mode $E_{IP}(x', y')$, the design wavelength, and the cavity dimensions, based on the desired image $I_{IP}(x', y')$.

identify the reflector type, e.g., metallic reflector or DBR (see section "DBR subtleties in metasurface microcavities" on favorable DBR characteristics) and the metasurface material.

calculate a metaatom library, i.e., the nanopillar diameter-dependent reflection phase for the chosen reflector and metasurface material. Adapt the transverse unit cell size and nanopillar height to achieve full phase coverage.

use equations (3) and (7) to calculate the metasurface phase $\phi_{MS}(x, y)$ from $E_{IP}(x', y')$.

discretize $\phi_{MS}(x, y)$ using the transverse unit cell dimensions and map the phase in each position to a pillar diameter to obtain the metasurface design.

**Numeric Validation**

After designing the metasurface phase profile and the metacavity, we validated the full cavity using finite-difference time-domain simulations (Flexcompute, Tidy3d, https://www.flexcompute.com/). We simulate the full metasurface cavity volume, circumscribed with perfectly matched layers in the propagation direction placed one wavelength before the entrance metallic partial reflector and one wavelength after the exit partial reflector of the microcavity.

In the simulation, we find the best representation of the skier in the image plane with a cavity length of ~37.0 μm, slightly shorter than our design length. This shift is caused by light penetration into the metasurface and the partial reflector, which changes the imaging condition. The effect has been previously observed for spherical microcavities[23,24]. Without a second partial reflector, i.e., without a cavity, the metasurface creates only a poor representation of the skier because the spatial intensity profile is uncontrolled.

This is verified by the simulation in Fig. 2e, which shows the intensity profile generated by the metasurface upon a single reflection. When adding the second partial reflector and closing the cavity, only the light that couples into the skier mode fulfills the round-trip condition. As a consequence, this light interferes constructively every round-trip, leading to field enhancement and intensity build-up in the skier mode (Fig. 2d).

Fig. 2b shows that the phase profile in the image plane is barely varying, in accordance with the spatially constant phase profile in the image plane that we chose during the design process. Fig. 2c shows the intensity profile of the fully built-up skier mode in the metasurface plane of the cavity, which has no similarity with the skier.

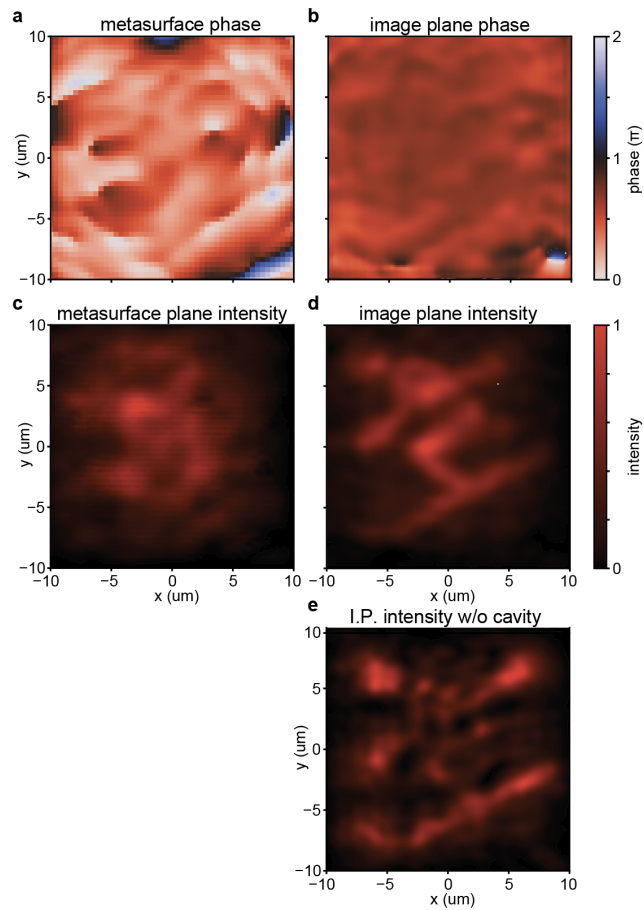

*Figure 2. Evolution from the spatial phase profile to the designed spatial intensity profile. a) Reflection phase of the metasurface. b) Modeled phase profile of the excited mode in the image plane. The simulation shows only minimal phase variation in the image plane. c) Intensity profile of the excited mode in the metasurface plane (modeled). In the metasurface plane, the excited mode does not resemble the incoming plane wave, nor does it resemble the designed skier intensity profile. d) In the image plane, the modeled excited mode intensity profile resembles the designed skier shape. e) The image quality strongly deteriorates when only reflection from the metasurface partial reflector is modeled without a second partial reflector to form the cavity.*

**Cavity Mirror and Metasurface Fabrication**

For the experimental realization of the cavity, we first fabricated two partial reflectors by depositing a chromium adhesion layer (5 nm) and a partially reflective gold film (50 nm) on fused silica substrates via electron-beam evaporation. Using plasma-enhanced chemical vapor deposition, we deposited a thin fused silica layer on one of the reflectors to aid the adhesion of the titania nanopillars to the gold. The titania metasurface was then fabricated on this sample following a previously presented approach[33]: the sample was coated with positive electron beam lithography resist (Zeon Corporation, ZEP-520A), baked, and covered with a conductive polymer layer (Showa Denko, ESPACER) to prevent charging effects. Electron-beam lithography was used to define a negative of the metasurface layout after development in o-xylene. This negative was then filled with titania using atomic layer deposition, creating the nanopillars. Lastly, excess titania on top of the resist was removed using reactive ion etching, and the resist was removed (MicroChem, Remover PG). Figs. 3c-e show scanning electron micrographs of the finished metasurface and close-ups of the nanopillar geometry.

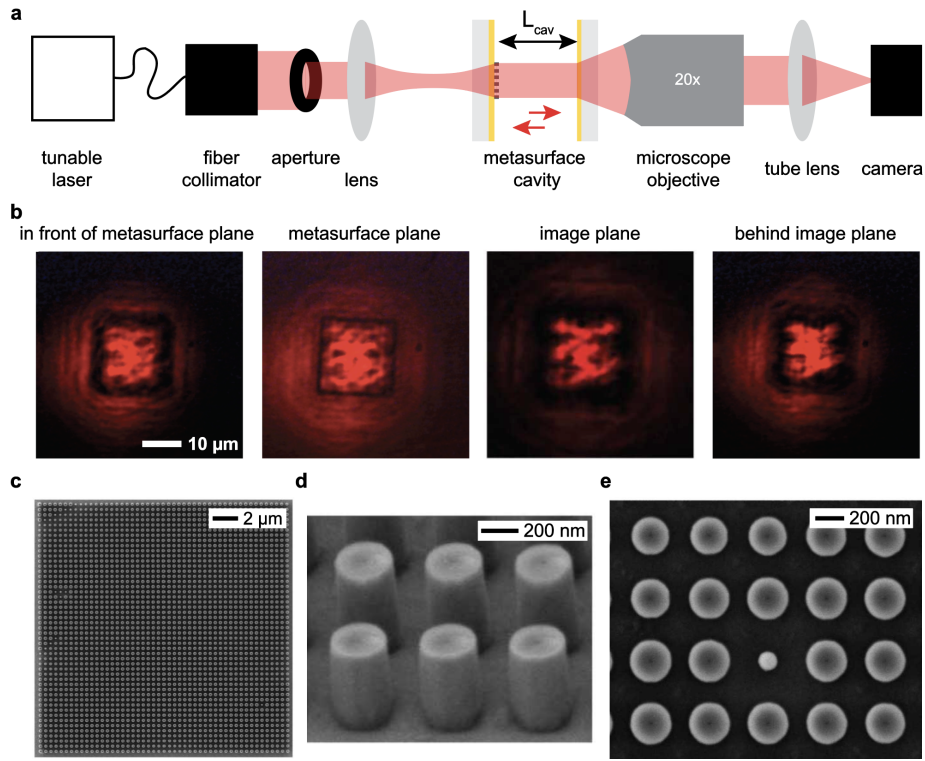

Figure 3. Experimental demonstration of a holographic microcavity. a) Optical measurement setup. Light from a tunable laser is coupled into the metasurface cavity by a lens. The beam size is reduced by an aperture. Using a microscope objective and a camera, we image the transverse mode profile for different cavity lengths (imaged by changing $L_{cav}$, controlled by a piezo stage, data in Fig. 4) and different transverse planes along the optical axis (imaged by shifting the focal plane of the microscope objective, data in panel b). Simultaneously, we measure the transmitted light's spectrum using a beam splitter between the objective and tube lens and a spectrometer (data in Fig. 4). The planes are along the optical axis, parallel to the metasurface, and are imaged by shifting the focus of the microscope objective through the cavity. b) Formation of the intracavity hologram. The four images show the transverse mode profile at different positions along the optical axis. During the measurement, the cavity length remained fixed. The four images depict the transverse mode profiles in: a plane ~40 µm in front of the microcavity, in the plane of the metasurface, in the image plane, and in a plane ~40 µm behind the cavity. It is apparent that the mode profile only resembles the skier in the image plane. The images were captured by scanning the microscope objective's focal plane along the optical axis. The scale bar applies to all four images. c) Scanning electron microscope (SEM) image of the entire metasurface. d) SEM side profile image of the nanopillars in the metasurface (400 nm in height). e) SEM image of a subsection of the metasurface, showing pillar diameter variation.

**Experimental Procedure**

Fig. 3a shows the experimental setup built to investigate the performance of the holographic microcavity. Light from a tunable laser (NKT Photonics, SuperK) is collimated into free space by a fiber collimator. Due to the small transverse size of the cavity, we weakly focus the incoming beam. To limit the resulting phase curvature of the incoming wavefront and uniformly illuminate the metasurface, we first reduce the size of the incoming beam to less than 0.5 mm using an aperture. Then, a focusing lens (focal length f = 3 cm) couples light into the cavity (resulting in a numerical aperture NA<0.01).

The metasurface reflector sits on a piezo-driven three-axis stage (Thorlabs, Nanomax) to allow movement relative to the laser beam and to tune the cavity length. We couple light to the cavity from the metasurface plane side to allow imaging of the cavity mode in the image plane with a 20x microscope objective, a tube lens, and a camera. At the same time, we record the spectrum of the cavity mode using a beam splitter and grating spectrometer (Andor, Shamrock, not shown).

## Results

**Experimental Results**

When a laser beam is incident on the cavity, and provided that the cavity is close to its design length, and the cavity is on resonance, we observe that the skier mode is excited. As seen in Fig. 3b, the shaped intensity profile builds up even though the cavity is illuminated by an unstructured beam. As expected, the skier intensity profile is only visible in the image plane, which we verified by scanning the microscope objective along the optical axis to image the mode profile at different longitudinal positions (Fig. 3b): the metasurface plane is recognizable by its square shape, however, in this plane, no skier is observed. The same is true for positions in front and after the microcavity.

We then focused on the length tuning of the cavity mode: we fixed the microscope objective position to image the image plane onto the camera sensor and swept the cavity length by changing the position of the metasurface reflector (compare with Fig. 3a). At each longitudinal position of the metasurface reflector, we recorded the transmitted spectrum with the laser tuned to $633 \pm 35$ nm (full width at half maximum intensity bandwidth). The broad bandwidth excites multiple longitudinal modes so we can deduce the optical cavity length (including the light's penetration in the metasurface and the reflectors) from the spectral spacing between these modes (the free spectral range $\lambda_{FSR}$). We then narrowed the incident laser bandwidth to 5 nm in order to excite only one longitudinal mode and recorded the spatial intensity distribution in the image plane and the transmitted spectrum.

Results are summarized in Fig. 4. We observe transmission maxima at cavity lengths spaced by $\lambda/2 = 316.5$ nm (see Fig. 4b). To judge the recreation of the designed mode, we calculate the normalized root-mean-square deviation at each resonance position, which is shown in Fig. 4c. We observe the best skier image at a cavity length of 40.8 μm (compare Fig. 4a and Fig. 4e), very close to the design length. In the spectral domain (see Fig. 4d), the mode exhibits a bandwidth of $\Gamma = 0.8$ nm, determined using a Lorentzian fit, which results in a quality factor of $Q \simeq 800$. The skier is also reproduced at other cavity lengths close to the design length, however, it starts deteriorating quickly and becomes indiscernible at more than ~5 μm away from the design length (Fig. 4e). Using the observed bandwidth and free spectral range ($\lambda_{FSR} = 4.8$ nm) at 40.8 μm cavity length, we find a round-trip loss of 64 %. This corresponds well to the predicted round-trip loss of 69 %, calculated from the partial reflector reflectance (46%) and the average metaatom reflectance used in the design (67%). This indicates that the intensity enhancement (2.6x) currently realized in the holographic metasurface microcavity is limited by our choice of mirror reflectance.

With mirror reflectance dominating the losses of the presented holographic cavity, smaller resonant linewidths, higher quality factors, and higher field enhancement in the cavity should be achievable by switching from metallic partial reflectors to DBRs[32]. These mirrors furthermore eliminate the coupling loss due to absorption in the metallic partial reflectors.

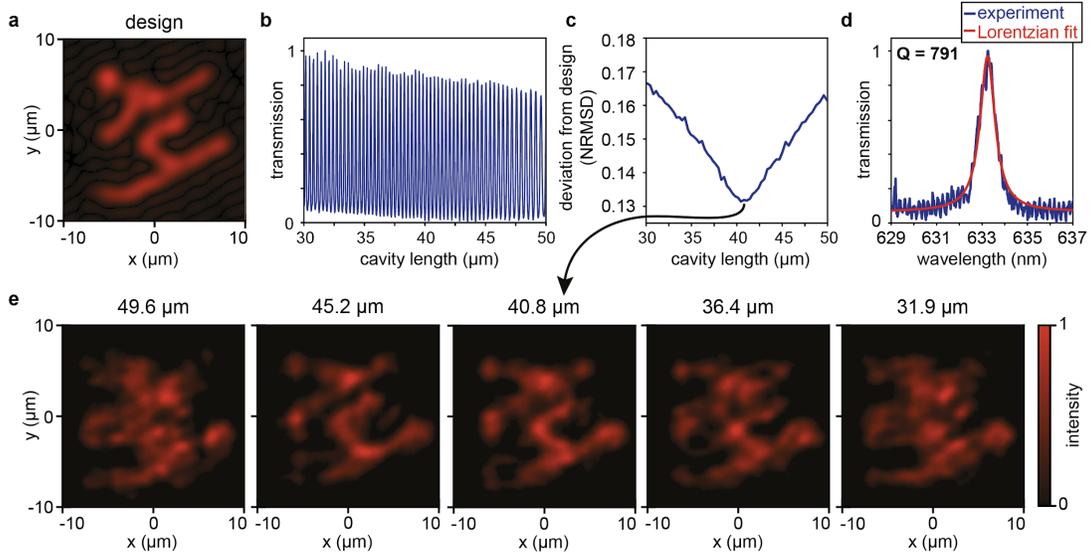

*Figure 4. Tuning of the microcavity mode. a) Spatial intensity profile of the target mode filtered such that it only contains transverse wavevectors that are supported by the finite cavity dimensions (20 μm x 20 μm x 40 μm, transverse, transverse, longitudinal). All false-color plots share the color bar in panel e. b) Cavity-length-dependent transmission of the cavity. c) Cavity-length-dependence of the normalized root-mean-square deviation (NRMSD) between the peak of the observed mode and that of the target mode. d) Measured spectrally resolved transmission (blue line) at 40.8 μm cavity length and a Lorentzian fit (red line) to the experimental data. e) Evolution of the spatial mode profile in the image plane when changing the cavity length. We experimentally observe the best-matching spatial cavity mode profile at 40.8 μm cavity length, close to the design length of 40.0 μm.*

**Increasing Performance Using DBRs**

To investigate the prospects of employing DBRs, we replace the gold layer in Fig. 1a with DBRs composed of four and five layer-pairs, each comprising a quarter-wave layer of a high-index-material (titania, $TiO_2$, $n_{TiO2} = 2.35$) and a quarter-wave layer of a low-index-material (fused silica, $SiO_2$, $n_{SiO2} = 1.46$), see Fig. 5a for an illustration. Results depend on the material of the DBRs' top layer (i.e., the material that touches the metasurface, see Fig. 5a). We will call a DBR terminated by fused silica a low-n capped DBR and by titania a high-n capped DBR, where n is the refractive index.

When placing nanopillars on these DBRs, narrowband resonances appear in the pillar-dependent reflectance when using the previous unit cell size and nanopillar height (because the reflector's materials are non-absorbing, long-lived resonances in the nanopillars are not damped, as they are by the gold mirror). To avoid these resonances, we decrease the unit cell size to 250 nm x 250 nm and increase the pillar height to 600 nm. We use a minimum pillar diameter of 70 nm and a minimum pillar spacing of 100 nm to maintain manufacturability[35]. The resulting nanopillar libraries offer complete phase coverage and a smooth phase and reflectance profile (see Figs. 5b and 5c). The number of layer pairs in the DBR controls the average reflectance. Still, it leaves the nanopillar-dependent reflection phase unchanged (see Fig. 5c). Changing the capping layer material introduces an expected π-phase shift[36].

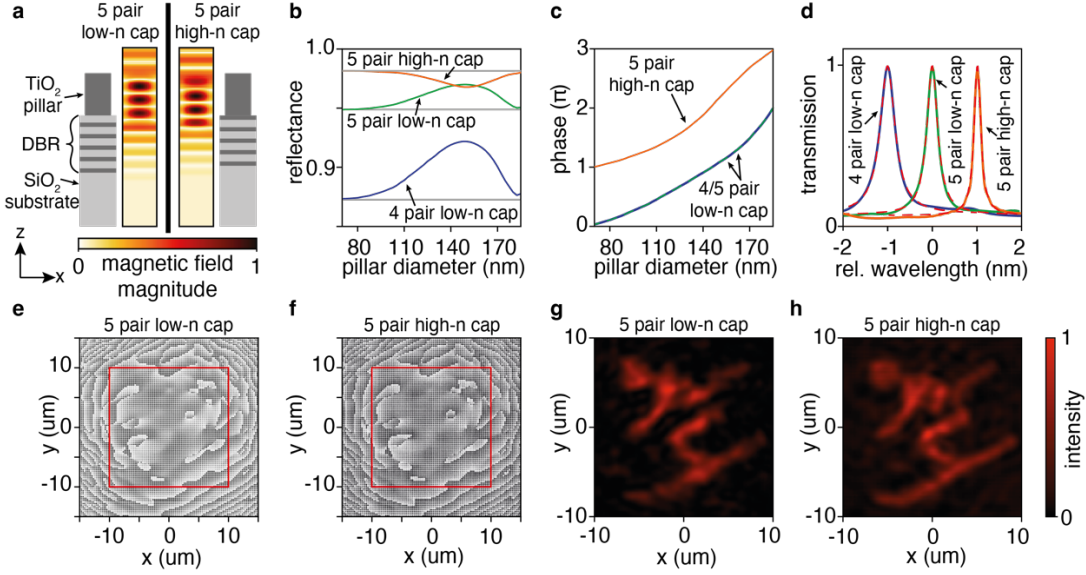

*Figure 5. Employing distributed Bragg reflectors (DBRs) in complex-shaped-mode metasurface microcavities. a) Side-view cross-sections of the new metasurface unit cells comprising a circular $TiO_2$ nanopillar (dark grey) on a $SiO_2/TiO_2$ DBR on a $SiO_2$ substrate (light grey). The DBR can be terminated by a low-n layer ($SiO_2$, left panel) or a high-n layer ($TiO_2$, right panel). Next to the schematics, false color plots show the magnetic field of a reflected light wave that penetrates deeper into the low-n-capped DBR (633 nm wavelength, 180 nm pillar diameter, 600 nm pillar height, 250 nm x 250 nm unit cell size). b) Diameter-dependent reflectance of nanopillars on different DBRs (colored lines). The grey lines mark the reflectance of the bare DBRs. c) Diameter-dependent reflection phase of nanopillars on different DBRs. The reflection phases of the low-n capped DBRs with four and five layers are indistinguishable. d) Transmission spectra (colored lines) of skier modes modeled by full-cavity simulations implementing the metasurfaces in panels e and f (designed using the libraries in panel c). The red dashed lines are least-squares fits of a Lorentzian function to the data. e) Metasurface design realizing a skier mode using a low-n capped DBR. The red square marks a 20 µm x 20 µm area. f) Metasurface design realizing a skier mode using a high-n capped DBR. g) Spatial cavity mode profile realized using a 30 µm x 30 µm metasurface on a low-n capped DBR (panel e). h) Spatial cavity mode profile realized using a 30 µm x 30 µm metasurface on a high-n capped DBR (panel f).*

Table 1: Full-cavity simulation results employing metasurfaces on different $SiO_2/TiO_2$ DBRs.

| DBR type | $R_{IP}$ | $R_{MP}$ | $L_{mirror}$ | metasurface size | spectral bandwidth of the skier mode | quality factor (x1000) | $L_{round-trip}$ | $L_{non-mirror}$ |
|---|---|---|---|---|---|---|---|---|
| 4-pair low-n cap | 87 % | 90 % | 22 % | 20 µm x 20 µm | 0.40 nm | 1.6 | 39 % | 22 % |
| 5-pair low-n cap | 95 % | 96 % | 9 % | 20 µm x 20 µm | 0.29 nm | 2.2 | 30 % | 23 % |
| 5-pair low-n cap | | | | 30 µm x 30 µm | 0.28 nm | 2.2 | 30 % | 23 % |
| 5-pair high-n cap | 98 % | 97 % | 5 % | 20 µm x 20 µm | 0.29 nm | 2.2 | 31 % | 24 % |
| 5-pair high-n cap | | | | 30 µm x 30 µm | 0.18 nm | 3.5 | 20 % | 16 % |

Picking from the three displayed libraries, we design metasurfaces to realize skier cavity modes (shown in Fig. 5e and 5f), implement them in complete cavities, and model their performance under plane-wave illumination using finite-difference time-domain simulations. We observe Lorentzian spectral mode profiles for all mirror configurations (see Fig. 5d). Using a 30 µm x 30 µm size metasurface placed on a high-n capped DBR, we achieve a designer-cavity-mode with a spectral linewidth of 0.18 nm, a quality factor larger than 3500, and a close reproduction of the target spatial profile (compare Fig. 5h with Fig. 4a).

Table 1 summarizes the bandwidths, quality factors, and round-trip losses[37] for cavities using different sizes and mirror configurations. The data reveals that the high-n capped DBRs perform better than their low-n capped counterparts. This observation is corroborated by the image quality (compare Fig. 5g and Fig. 5h): although all cavities create a skier mode, the high-n capped DBR cavity's rendition of the skier shows less speckle and its outline is truer to the design image.

**DBR Subtleties in Metasurface Microcavities**

To understand this difference, we explore the different loss mechanisms in a metasurface microcavity: Light is lost when it transmits through the cavity end mirrors. Our cavities comprise a metasurface-covered DBR (reflectance: $R_{MP}$) and an opposing DBR (reflectance: $R_{IP}$), yielding a total mirror loss per round-trip $L_{mirror} = 1 - R_{IP}R_{MP}$. Light is also lost when it diffracts outside the cavity (diffraction loss $L_{diffr}$) due to insufficient mirror size. Employing a larger metasurface mitigates this loss mechanism. Furthermore, light can be scattered outside the cavity, e.g., by phase errors introduced by the metasurface (scattering loss $L_{scat}$). Combined, the round-trip loss is

$$L_{round-trip} = 1 - (1 - L_{mirror})(1 - L_{diffr})(1 - L_{scat}) = 1 - (1 - L_{mirror})(1 - L_{non-mirror}) \qquad (8)$$

where, on the second line, the non-mirror loss $L_{non-mirror}$ comprises the diffraction and scattering losses.

The results in Table 1 reveal that for all 20 μm x 20 μm large metasurfaces, the non-mirror losses are approximately 23%. Increasing the metasurface size to 30 μm x 30 μm should decrease the diffraction loss from 13% to 2% (and thus the non-mirror loss to approximately 14%) and improve performance. Interestingly, the performance only increases for the high-n capped cavity and remains unchanged for the low-n capped cavity. Because the metasurface designs are alike (compare Figs. 5e and 5f), the root cause must be the DBRs: light penetrates deeper into a low-n capped DBR than into a high-n capped DBR[36,38]. This fact holds when a metasurface is on the DBR, visible in the field distributions plotted in Fig. 5a.

When increasing the metasurface size, we add regions (see the areas outside of the red squares in Fig. 5e and 5f) that require steep phase profiles (much like the outside parts of a lens with a high numerical aperture). In these regions, neighboring nanopillars must introduce larger phase differences than in the center of the metasurface and are less similar. When the light penetrates the DBR, it is unconfined (unlike in the metasurface's titania nanopillars) and is more prone to mix with light from the neighboring unit cells. Consequently, due to their lower light penetration, high-n capped DBRs allow less mixing between the light fields of adjacent metasurface unit cells and can implement higher phase slopes, explaining their better performance and image quality. This finding extends beyond implementing microcavities and should also be taken into consideration when designing other metasurface-DBR hybrid reflectors.

For small spectral bandwidths, which are common in cavity applications, and simple mode profiles, scattering losses at the percent level have been realized[24]. The rapid progress in novel design techniques and optical metasurfaces[39] promises similar efficiencies for complex-shaped mode profiles and broader bandwidths in the near future.

# CONCLUSION

Previous microcavities have focused on simple mode shapes due to fabrication limitations. Free-form mode profiles have so far relied on injecting an image and macroscopic setups to achieve imaging functionality[11,12,13]. As shown, the holographic metasurface microcavity uses the design flexibility and miniature scale of optical metasurfaces to fulfill the round-trip condition for a complex-shaped mode profile.

The presented design mechanism can realize arbitrary and asymmetric mode profiles within physical limits: First, the minimal achievable feature size in the image plane is given by the size of the metasurface and is limited by the Abbe diffraction limit. This can effectively be overcome as metasurfaces today can be manufactured with centimeter scale and larger transverse dimensions[40,41] and at ultraviolet frequencies[42,43,44,45]. If a use case requires changing conserved quantities of the incoming light, e.g., its orbital angular momentum (OAM)[46], two different metasurface partial reflectors are required to maintain an intact round-trip phase condition.

We foresee that holographic metasurface microcavities can be useful for various applications.

Semiconductor lasers, including quantum dot lasers[47], quantum-cascade lasers[48], and vertical-cavity surface-emitting lasers, have non-spherical intracavity modes and output modes due to their geometry. Holographic microcavities can optimally illuminate the non-spherical gain materials in such lasers by shaping their intracavity spatial mode profile. They can also efficiently couple such lasers to external microcavities or provide spatially tailored feedback, e.g., to balance the thermal load in such lasers.

Optical lattices are established by standing wave patterns formed by counter-propagating light beams. Cavities enhance such standing waves[49]. Currently, modes with large diameters are employed to increase the intensity uniformity at their center. When using two metasurface reflectors, our approach can create an image plane in the free space between them. A cavity hologram can then increase the mode uniformity to miniaturize the optical lattice assembly, or to shape the optical lattice.

Microscale intracavity image enhancement could also potentiate optical signal processing, as has been previously suggested[13]. By adding a thin or two-dimensional nonlinear material in the image plane of the metacavity and tailoring the holographic mode to couple spatially separate spots of the nonlinear material, optical thresholding depending on multiple input parameters[50] could be realized. This capability would be helpful, e.g., for optical artificial neural networks.

Furthermore, cavity-enhanced microscopy promises to enhance detection sensitivity via multiple light-matter interactions[6] or the Purcell effect[51]. Holographic microcavities could combine sensitivity enhancement with structured illumination by light sheets or Bessel beams.

Our solution uses widespread and industrially scalable fabrication protocols, and thus can directly be adopted at an industry scale.

# AUTHOR INFORMATION

## CORRESPONDING AUTHOR

* mossiander@g.harvard.edu

Present Addresses

† E.L. Ginzton Laboratory, Stanford University, 94305 Stanford, California, USA

‡ Institute of Nanotechnology, Karlsruhe Institute of Technology (KIT), 76344 Eggenstein-Leopoldshafen, Germany

## AUTHOR CONTRIBUTIONS

S.M. and M.O. developed the project. S.M. designed the metasurface. M.L.M. fabricated the samples. S.M. imaged the samples. S.M., C.S., and M.O. built the experimental setup and experimented. S.M. and M.O. performed numerical modeling. S.M., M.O., and F.C. wrote the manuscript.

## ACKNOWLEDGMENT

The authors acknowledge a Tidy3d computation power gift from Zongfu Yu at Flexcompute Inc.

## FUNDING SOURCES

This work was performed, in part, at the Center for Nanoscale Systems (CNS), a member of the National Nanotechnology Coordinated Infrastructure (NNCI), which is supported by the NSF under award no. ECCS-2025158. CNS is a part of Harvard University. The authors acknowledge partial financial support from the Air Force Office of Scientific Research (AFOSR), under award number FA9550-21-1-0312. M.O. acknowledges funding by the Alexander von Humboldt Foundation (Feodor-Lynen Fellowship), by the Austrian Science Fund (FWF) Start Grant Y1525, and by the European Union (grant agreement 101076933 EUVORAM). For the purpose of open access, the authors have applied a CC BY public copyright license to any Author Accepted Manuscript version arising from this submission. Views and opinions expressed are however those of the author(s) only and do not necessarily reflect those of the European Union or the European Research Council Executive Agency. Neither the European Union nor the granting authority can be held responsible for them.